# Correlated pressure effects on structure and superconductivity in $LaFeAsO_{0.9}F_{0.1}$

**G. Garbarino**
European Synchrotron Radiation Facility (ESRF), 6 Rue Jules Horowitz 38043 BP 220 Grenoble Cedex

**P. Toulemonde, M. Álvarez-Murga, A. Sow**
Institut NEEL, CNRS & Université Joseph Fourier BP166, 25 Avenue des Martyrs, F-38042 Grenoble Cedex 9 France

**M. Mezouar**
European Synchrotron Radiation Facility (ESRF), 6 Rue Jules Horowitz 38043 BP 220 Grenoble Cedex

**and M. Núñez-Regueiro**
Institut NEEL, CNRS & Université Joseph Fourier BP166, 25 Avenue des Martyrs, F-38042 Grenoble Cedex 9 France

[click here and begin typing abstract] We have studied the structural and superconductivity properties of the compound $LaFeAsO_{0.9}F_{0.1}$ under pressures up to $32 GPa$ using synchrotron radiation and diamond anvil cells. We obtain an ambient pressure bulk modulus $K_0 = 78(2) GPa$, compressibility comparable to some cuprates. At high pressures, the sample is in the overdoped region, with a linear decrease with pressure variation of the superconducting transition temperature.

email: nunez@grenoble.cnrs.fr

72.80.Ga, 61.50.Ks, 62.50.+p

The discovery of superconductivity[1, 2, 3, 4] with critical temperatures between 30 and 55 K in two families of layered iron arsenides has triggered a large amount of work on the subject, as the strong magnetic character of these oxypnictides suggest the possibility of unconventional superconductivity mechanisms. Pressure measurements are of great utility in the study of superconductivity, as is exemplified by the highest measured superconducting[5] transition temperature in fluorinated $HgBa_2Ca_2Cu_3O_{8+\delta}$, $T_c = 166K$ at 26GPa, or the superconductivity in the high $\varepsilon$ pressure phase of iron[6]. Coupling the variation under pressure of superconducting properties and the lattice of the studied compound can be key in the comprehension of the materials, and are extremely useful to be compared with theoretical calculations. In particular, the up to now measured pressure properties of the $LaFeAsO_{1-x}F_x$ system presents some similarities with the behavior of cuprates (for a review see Ref. 7), in particular the passage under pressure from an underdoped to overdoped regime[2], implying a charge transfer under pressure between the $LaO$ and $FeAs$ layers. In order to provide elements for this type of analysis we have performed measurements of the evolution of the structure and of the superconducting properties of similarly prepared samples of $LaFeAsO_{0.9}F_{0.1}$.

Our samples $LaFeAs(O_{1-x}F_x)$ were prepared under high pressure – high temperature. Different mixtures of reactants were used. All the preparation and assembly of the high pressure cell were performed in a glove box filled with pure argon. For $x = 0$ powders of $Fe$ (Prolabo 99.5%), $Fe_2O_3$ (StremChem 99.8%), $As$ (Johnson Matthey Chemicals) were mixed with very small pieces of pure $La$ (Aldrich 99%). For the fluorine doped samples with $x = 0.1$, nominal mixtures of $FeAs$ (CERAC 99.5%), $La_2O_3$, $LaF_3$ (Merck) and $La$ or $Fe$, $Fe_2O_3$, $As$ and $LaAs$ (prepared as in Ref. 8) were used. Each mixture was pressed into a pellet and introduced in closed home made $h-BN$ crucibles which are placed into a tubular carbon furnaces. The whole assembly was put in the high pressure gasket made of pyrophillite. The setup was pressurized at $3-3.5 GPa$ in a belt type apparatus and heated at 1200°C for a dwell time of 1 to 4 hours, then quenched to room temperature. The conditions were optimized to decrease the proportion a $FeAs$, $LaAs$ and $LaOF$ impurities and obtain nearly pure phases (quickly converted in $La(OH)_3$ in contact with air).

The angle dispersive X-ray diffraction studies on $LaFeAsO_{0.9}F_{0.1}$ powder samples were performed at the ID27 high-pressure beamline of the European Synchrotron Radiation Facility using monochromatic radiation (λ=0.3738Å) and diamond anvil cells with 350µm cullet diamonds. Two transmitting media were used, 4:1 Methanol-Ethanol mixture for the low pressure range ($P < 3GPa$), and Ne for the high pressure one ($P > 3GPa$). The pressure was determined using the shift of the fluorescence line of the ruby. All the structural studies have been done at ambient temperature. The diffraction patterns were collected with a CCD camera, and the intensity vs. 2Theta patterns were obtained using the fit2d software[9]. A complete Rietveld refinement was done with the GSAS-EXPGUI package[10].

The electrical resistance measurements were performed using a Keithley 238 source meter and a Keithley 2182 nanovoltmeter. Pressure measurements, $1.4-22GPa$ (between 4.2K and 300K), were done in a sintered diamond Bridgman anvil apparatus using a pyrophillite gasket and two steatite disks as the pressure medium[11].

On Fig.1(a) we show the ambient pressure X-ray diffraction pattern and the Rietveld refinement. The lattice parameters are $a = 4.0040(1)$Å and $c = 8.6898(4)$Å for the

$LaFeAsO_{0.9}F_{0.1}$ sample used in the high pressure XRD experiment. Compared to the values for pure $LaFeAsO$ [12] and fluorine doped samples from the literature[1], our sample state corresponds to an overdoped sample. Unfortunately, by x-ray diffraction, it is not possible to determine the real fluorine content of the sample by Rietveld refinement. Nevertheless, the occupancy factor of the (O,F) site was estimated to be nearly full and probably the fluorine content is near the nominal composition, i.e. x= 0.1, because the HP-HT treatment is made in closed conditions and the loss of fluorine should be negligible.

The pressure evolution of the X-ray diffraction patterns can be seen on Fig.1(b). No structural transition is observed up to the highest measured pressure, $32 GPa$. From the Rietveld refinements, we obtain the pressure dependence of the $z$ atomic position of $La$ and $As$ (see Fig.1(c)) that shows a weak increment in the range $P < 10 GPa$, and then a saturation, with no signature of a phase transition.

The unit cell volume, $V$, at various pressures P, were, fitted to a third order Murnaghan equation of state $V = V_0 \left(1 + K_0' P / K_0 \right)^{-1/K_0}$, where $K_0$ is the bulk modulus at ambient conditions and $K_0' = 7.4(2)$. We obtain a value for the bulk modulus of $78(2) GPa$, that is very similar to those found in cuprates[13], but slightly lower than the $98 GPa$ obtained from theoretical calculations[14]. In Table I we show the values of the lattice parameters of $LaFeAsO_{0.9}F_{0.1}$ for the different measured pressures.

It is interesting to note that all the superconducting compounds with the highest $T_c$ (cuprates, $MgB_2$, oxypnictides) are of layered structure. As such, it is important to determine how pressure changes the interaction between the layers, interaction that can be measured by the relative compression of the $c$ parameter with respect to the $a$ parameter. We observe in Fig. 2(b) that it is more important in the shown cuprate and in the $Na_{0.5}CoO_2$ cobaltite than in $LaFeAsO_{0.9}F_{0.1}$, implying a less two-dimensional character in the oxypnictides, at least in what considers lattice properties.

On Fig. 3 we show the evolution of the temperature dependence of resistance of $LaFeAsO_{0.9}F_{0.1}$ as a function of pressure. The absolute value decreases with compression, as well as the resistance slope $\alpha$. The resistivity of a metal can be written[15]

$$R = \frac{12\pi^3\hbar}{e^2 \int_{FS} \Lambda_k dS_F} \approx \frac{12\pi^3\hbar}{e^2 \Lambda} \frac{1}{\int_{FS} dS_F}$$

Where $e$ is the electronic charge, and $\Lambda_k$ the mean free path for each vector $k$ on the Fermi surface (FS). If we approximate $\Lambda_k \approx \Lambda$, constant on all the FS, the inverse of the resistance slope, $\alpha^{-1}$, is directly proportional to the area of the FS. We observe that this parameter increases linearly with pressure, that can be interpreted, in a first approximation, as a constant charge transfer $dn/dP$ with pressure, as has been used to describe the behavior of cuprates under pressure7. Hall constant measurements under pressure would be necessary to confirm this assumption.

On Fig. 4(a) we show the variation of the superconducting transition temperature of our sample compared to the one reported by Takahashi et al[2]. $T_c$ onset is defined as in ref. 2, while $T_c mid$ is obtained from the peak in the derivative of the resistance. The comparison suggests that our sample in more on the overdoped region than the one measured by Takahashi et al., although the nominal composition is similar. It should be noted that the samples used for pressure measurements are extremely small, and that weak non-homogeneities in the bulk sample can result in having a measured that is not of the nominal composition. In any way, we observe a linear variation of $T_c$, that would imply, if we accept the constant charge transfer, that the dependence of

$T_c$ with carrier concentration is not strictly parabolic as in cuprates or that other factors come into play. According to Levy and Olsen[16], within conventional electron-phonon coupling theory BCS theory the logarithmic volume dependence of $T_c$ follows

$$\frac{d\ln(T_c/\Theta_D)}{d\ln V} = \ln\left(\frac{\Theta_D}{T_c}\right)\frac{d\ln(\lambda)}{d\ln V} \equiv \ln\left(\frac{\Theta_D}{T_c}\right)\varphi$$

where $\Theta_D$ is the Debye temperature, $\lambda$ the electron-phonon coupling parameter and $\varphi \approx 2.5$ for conventional superconductors. From our results on our variation of $T_c$ and volume with pressure on $LaFeAsO_{0.9}F_{0.1}$ and using[17] a $\Theta_D = 316K$, we obtain a $d\ln(T_c/\Theta_D)/d\ln V = 6.5$, both for the onset and the middle of the transition, see Fig. 4(b). We obtain then $\varphi \approx 2.75$, compatible with electron-phonon coupling, although on one hand, Levy and Olsen's semi-empirical analysis is valid at low pressures ($<1GPa$) and must be correlated with a hypothetic isotope effect to be conclusive. On the other hand we have ignored any variation of $\Theta_D$ with pressure, as no measurement is available.

As we are seemingly in an overdoped sample, we can look for Fermi liquid behavior as has been observed in cuprates, in electrical resistance a quadratic term in temperature due to Landau quasiparticle-quasiparticle scattering. The resistance above the superconducting transition follows approximately a law $A \cdot T^n$, with $n \approx 2$, see Fig.4(c). However, we must strongly remark that this is not the signature of a Fermi liquid behavior, as $A$ follows a linear law with the residual resistance $R_0$ as pressure is changed, with a coefficient of $6 \cdot 10^{-5}$, Fig. 4 (d), indicating conclusively that carrier scattering preceding the superconducting transition in the sample is dominated by inelastic scattering against defects and impurities, the Koshino-Taylor mechanism[18].

Samples with less defects and impurties will be needed to determine if the overdoped region is dominated by Fermi liquid quasiparticle-quasiparticle scattering as in cuprates. In conclusion, we have measured the structural and transport properties of $LaFeAsO_{0.9}F_{0.1}$. We find no evidence of a phase transition up to the highest pressure studied and a compressibility similar to that of cuprates and an apparent constant charge transfer under pressure.

We thank Murielle Legendre for the preparation of the home made $h-BN$ crucibles, and W. Crichton and J.P. Perrillat for help in the diffraction measurements.

| Pressure (GPa) | a (Å) | c (Å) | Volume (Å$^3$) |
| --- | --- | --- | --- |
| 0.00 | 4.0040(1) | 8.6898(4) | 139.31(1) |
| 0.82 | 3.9986(1) | 8.6410(3) | 138.16(1) |
| 3.16 | 3.9705(1) | 8.5477(3) | 134.75(1) |
| 4.58 | 3.9529(2) | 8.4678(7) | 132.32(1) |
| 4.85 | 3.9514(2) | 8.4574(6) | 132.05(1) |
| 5.68 | 3.9443(2) | 8.4325(6) | 131.19(1) |
| 6.36 | 3.9387(2) | 8.4137(7) | 130.52(1) |
| 7.08 | 3.9329(2) | 8.3930(6) | 129.82(1) |
| 8.06 | 3.9251(2) | 8.3646(7) | 128.87(1) |
| 8.60 | 3.9216(2) | 8.3476(8) | 128.38(1) |
| 9.19 | 3.9184(2) | 8.3327(8) | 127.94(1) |
| 9.90 | 3.9143(3) | 8.3116(9) | 127.35(1) |
| 10.91 | 3.9085(3) | 8.2834(9) | 126.54(1) |
| 12.22 | 3.9012(3) | 8.2549(10) | 125.62(1) |
| 13.41 | 3.8956(3) | 8.2247(10) | 124.82(1) |
| 15.22 | 3.8869(3) | 8.1833(11) | 123.64(2) |
| 16.70 | 3.8793(4) | 8.1513(12) | 122.67(2) |
| 18.10 | 3.8733(4) | 8.1214(13) | 121.84(2) |
| 19.56 | 3.8671(4) | 8.0913(14) | 121.00(2) |
| 21.46 | 3.8603(5) | 8.0513(15) | 119.98(2) |
| 23.69 | 3.8519(5) | 8.0089(16) | 118.83(3) |
| 26.63 | 3.8436(6) | 7.9493(19) | 117.43(3) |
| Pressure (GPa) | a (Å) | c (Å) | Volume (Å$^3$) |
| 30.87 | 3.8345(7) | 7.8494(25) | 115.41(3) |
| 32.26 | 3.8310(8) | 7.8330(30) | 114.96(4) |

Table I. Measured lattice parameters of as a function of applied pressure. The error in pressure determination is estimated to 0.05GPa.

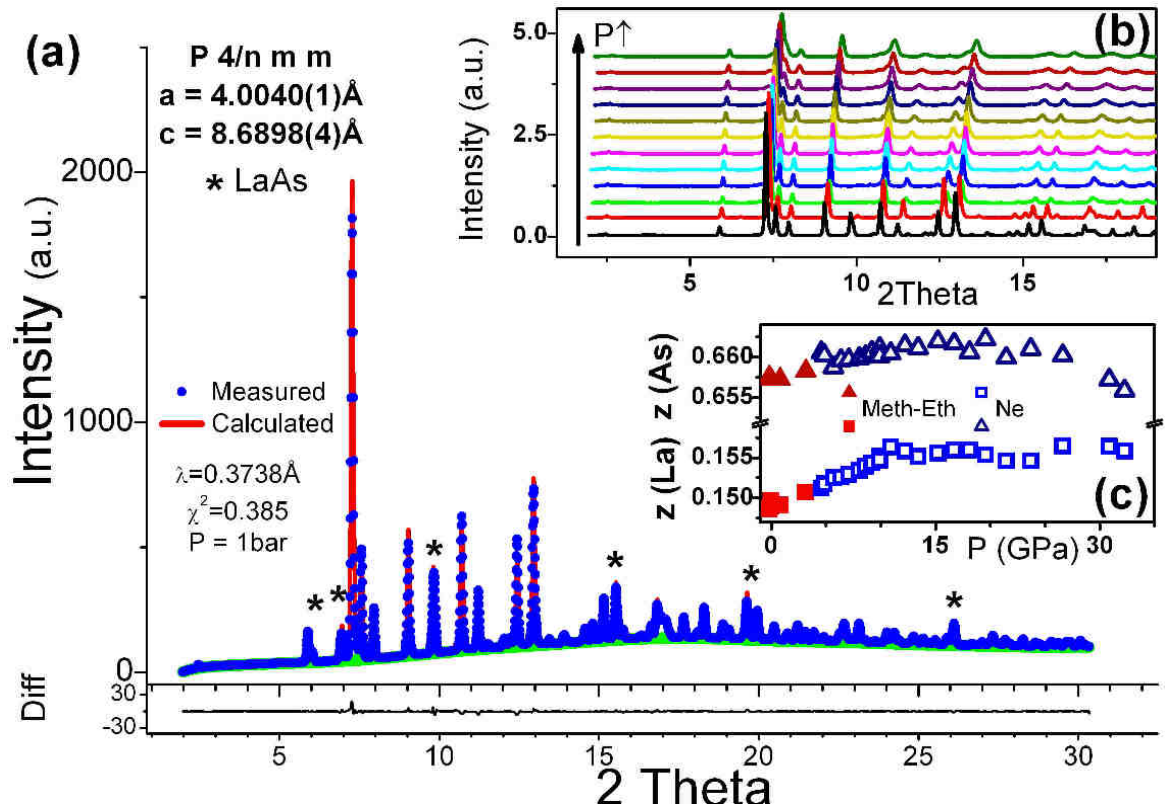

**Figure1: (color online) (a)** X-ray synchrotron radiation diffraction pattern of the $LaFeAsO_{0.9}F_{0.1}$ powder sample at ambient pressure. The Rietveld refinement is the red solid line. The black stars represent the $LaAs$ impurity phase.

**(b)** Pressure evolution of the diffraction patterns of $LaFeAsO_{0.9}F_{0.1}$. The solid arrow indicate the increasing pressure sense. The data correspond to 0; 3.2; 4.6; 5.7; 7.1; 8.6; 9.9; 12.2; 15.2; 18.1; 21.5 and 30.9GPa, respectively.

**(c)** Pressure dependence of the $z$ atomic position of the $La$ and $As$ atoms. Solid symbols correspond to Methanol-Ethanol pressure media, while open symbols to Ne media.

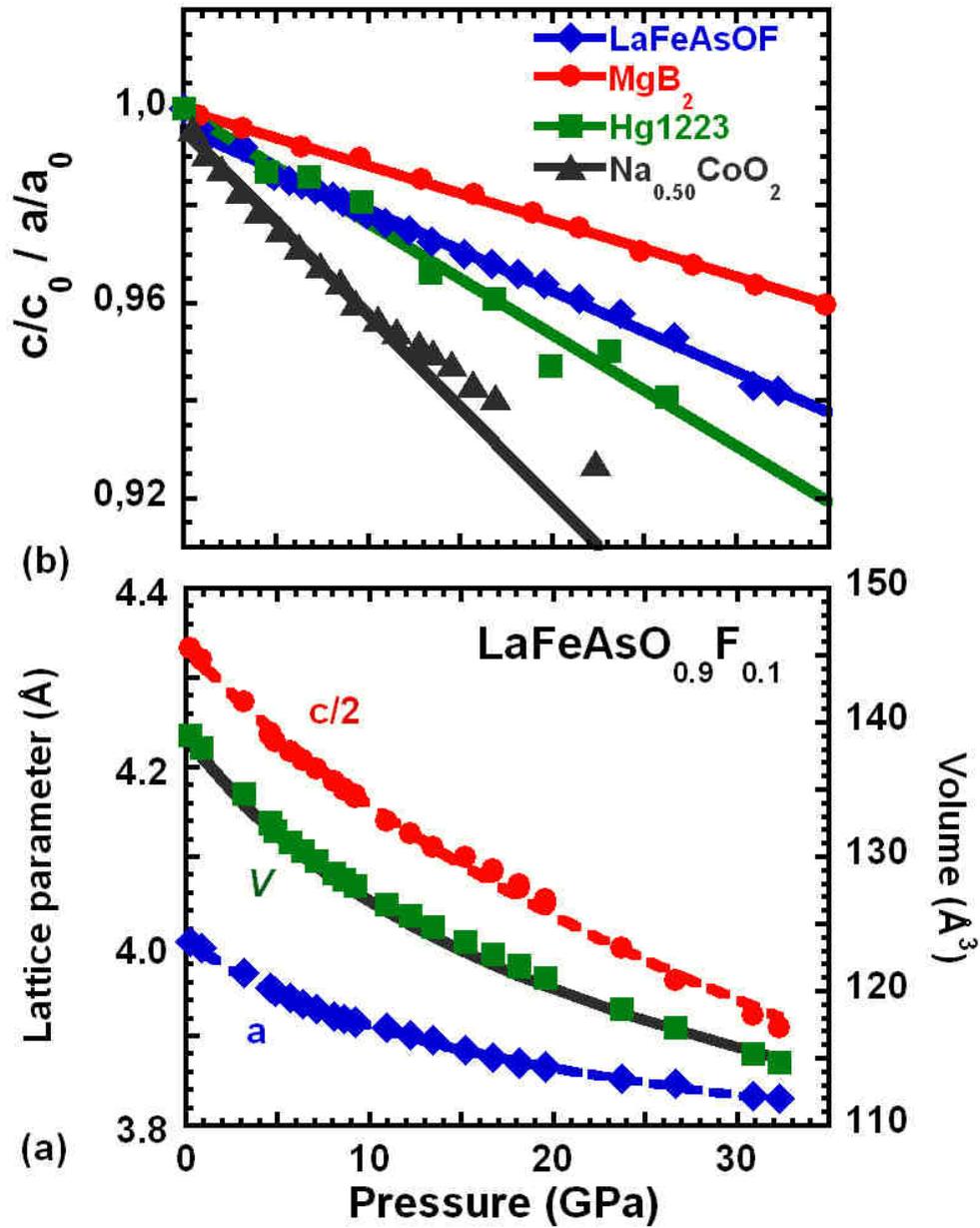

**Figure 2:** (color online) (a) Evolution of the lattice parameters of as a function of pressure. Blue diamonds: $a$ ; red dots: $c/2$ ; green squares : volume. The solid black line corresponds to Murnaghan equation of state (see text). (b) Comparison of the ratio of the basal parameter to the ratio of the stacking parameter for different high temperature superconductors. Blue diamonds: $LaFeAsO_{0.9}F_{0.1}$ (this paper); red dots: $MgB_2$[19]; green squares: $Hg-1223$[13].

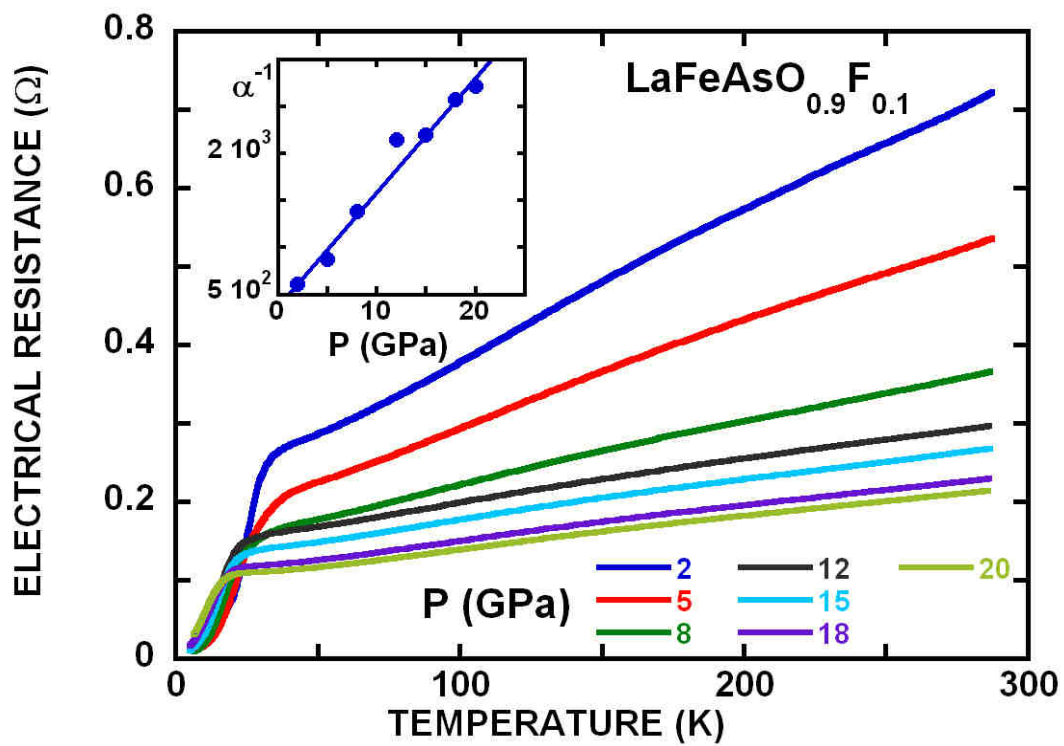

**Figure 3: (Color online)** Electrical resistance of $LaFeAsO_{0.9}F_{0.1}$ sample as a function of temperature for different pressures a indicated in the figure.

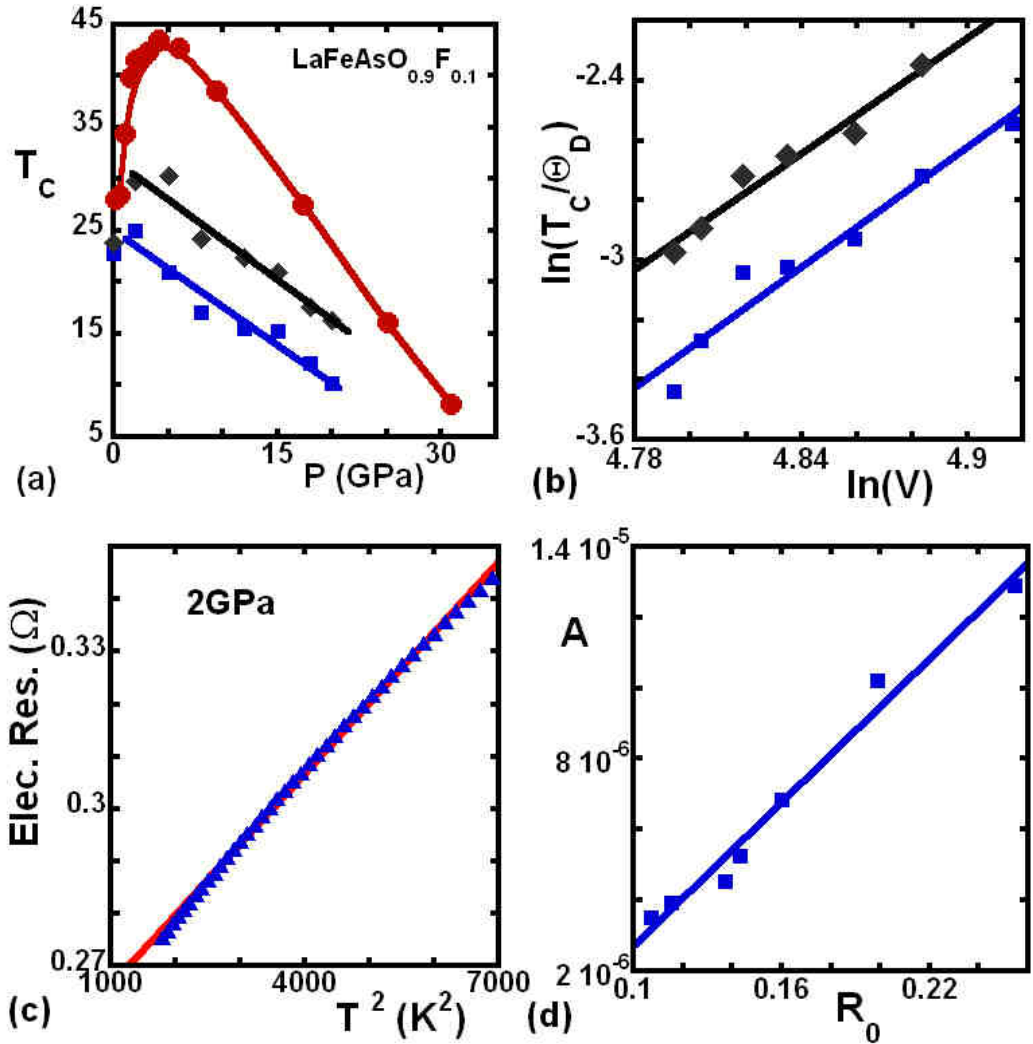

**Figure 4:** (color online) (a) Superconducting $T_c$'s as a function of pressure : red dots Takahashi et al. (Ref. 2) onset data ; black diamonds: our onset data; blue squares : our mid transition data. (b) Dependence of the logarithmic ratio of Tc with the logarithm of the volume as a function of pressure. (c) Electrical resistance of the sample at 2GPa as a function of the square of the temperature showing the $R = R_0 + A \cdot T^2$ law. (d) Linear dependence of $A$ with $R_0$. The value of the slope, $6.10^{-5}$, implies that the quadratic dependence is due to the Koshino-Taylor mechanism, inelastic scattering against impurities.